\documentclass[twocolumn, apj]{emulateapj}
\usepackage{amsmath, epsfig}
 
\newcommand{\dif}{\mathrm{d}}
\newcommand{\degree}{\ensuremath{^\circ}}
\def\la{\langle}
\def\ra{\rangle}

\shorttitle{Heat Flux Driven Buoyancy Instability}
\shortauthors{Parrish \& Quataert}

\begin{document}

\title{Nonlinear Simulations of the Heat Flux Driven Buoyancy
Instability and its Implications for Galaxy Clusters}

\author{Ian J. Parrish\altaffilmark{1, 2} and Eliot Quataert\altaffilmark{1}}

\altaffiltext{1}{Astronomy Department \& Theoretical Astrophysics Center, 601 Campbell Hall, The University of California, Berkeley, CA 94720; iparrish@astro.berkeley.edu; eliot@astro.berkeley.edu}
\altaffiltext{2}{Chandra Fellow}

\begin{abstract}

In low collisionality plasmas heat flows almost exclusively along
magnetic field lines, and the condition for stability to convection is
modified from the standard Schwarzschild criterion.  We present local
two and three-dimensional simulations of a new heat flux driven
buoyancy instability (the HBI) that occurs when the temperature in a
plasma decreases in the direction of gravity.  We find that the HBI
drives a convective dynamo that amplifies an initially weak magnetic
field by a factor of $\sim 20$.  In simulations that begin with the
magnetic field aligned with the temperature gradient, the HBI
saturates by rearranging the magnetic field lines to be almost purely
perpendicular to the initial temperature gradient.  This magnetic
field reorientation results in a net heat flux through the plasma that
is less than 1\% of the field-free (Spitzer) value.  We show that the
HBI is likely to be present in the cool cores of clusters of galaxies
between $\sim 0.1-100$ kpc, where the temperature increases outwards.
The saturated state of the HBI suggests that inward thermal conduction
from large radii in clusters is unlikely to solve the cooling flow
problem.  Finally, we also suggest that the HBI may contribute to
suppressing conduction across cold fronts in galaxy clusters.
\end{abstract}

\section{Introduction} \label{sec:intro}
\vspace{-0.07in} For thermally stratified fluids, stability to
convection is guaranteed if the entropy decreases in the direction of
gravity.  In dilute astrophysical plasmas, however, the mean free path
is large compared to the gyroradius, and heat conduction is
anisotropic with respect to the magnetic field \citep{brag65}.  In the
low collisionality regime, the convective instability condition is
modified, and plasmas in which the temperature increases in the
direction of gravity are buoyantly unstable \citep{bal00}.  This
instability, called the magnetothermal instability (MTI), has been
simulated in two and three dimensions by \citet{ps05, ps07}.

In this {\it Letter}, we study a related instability of low
collisionality plasmas.  \citet{q07} has shown that, in the presence of
a background heat flux, anisotropic conduction drives a buoyancy
instability when the temperature decreases in the direction of
gravity.  The growth rates of this heat flux driven buoyancy
instability (HBI) are
verified using two-dimensional (2D) simulations in \S\ref{sec:2D}.  In
\S\ref{sec:3D} we determine the nonlinear saturation of the HBI in
three dimensions (3D) and show that it drives a magnetic dynamo while
greatly suppressing the net heat flux through the plasma.  Finally, in
\S\ref{sec:conc} we
apply the HBI to the intracluster medium and discuss its implications
for cooling flows and cold fronts.
\vspace{-0.11in}
\section{2D Simulations} \label{sec:2D}
\vspace{-0.07in}
\subsection{Equations and Method}\label{subsec:MHD}
\vspace{-0.07in}
We solve the usual equations of magnetohydrodynamics (MHD) with the
addition of anisotropic thermal conduction. The MHD equations in
conservative form are
\begin{equation}
\frac{\partial \rho}{\partial t} + \boldsymbol{\nabla}\cdot\left(\rho \boldsymbol{ v}\right) = 0,
\label{eqn:MHD_continuity}
\end{equation}
\begin{equation}
\frac{\partial(\rho\boldsymbol{v})}{\partial t} + \boldsymbol{\nabla}\cdot\left[\rho\boldsymbol{vv}+\left(p+\frac{B^{2}}{8\pi}\right)\mathbf{I} -\frac{\boldsymbol{BB}}{4\pi}\right] + \rho\boldsymbol{g}=0,
\label{eqn:MHD_momentum}
\end{equation}
\begin{equation}
\frac{\partial E}{\partial t} + \boldsymbol{\nabla}\cdot\left[\boldsymbol{v}\left(E+p+\frac{B^{2}}{8\pi}\right) - \frac{\boldsymbol{B}\left(\boldsymbol{B}\cdot\boldsymbol{v}\right)}{4\pi}\right] 
+\boldsymbol{\nabla}\cdot\boldsymbol{Q} +\rho\boldsymbol{g}\cdot\boldsymbol{v}=0,
\label{eqn:MHD_energy}
\end{equation}
\begin{equation}
\frac{\partial\boldsymbol{B}}{\partial t} + \boldsymbol{\nabla}\times\left(\boldsymbol{v}\times\boldsymbol{B}\right)=0,
\label{eqn:MHD_induction}
\end{equation}
where the symbols have their usual meaning. The total energy $E$ is given by
\begin{equation}
E=\epsilon+\rho\frac{\boldsymbol{v}\cdot\boldsymbol{v}}{2} + \frac{\boldsymbol{B}\cdot\boldsymbol{B}}{8\pi},
\label{eqn:MHD_Edef}
\end{equation}
where $\epsilon=p/(\gamma-1)$.  Throughout this paper, we assume
$\gamma=5/3$.  The anisotropic electron heat flux is given by
\begin{equation}
\boldsymbol{Q} = - \chi_{C} \boldsymbol{\hat{b}\hat{b}}\cdot\boldsymbol{\nabla}T,
\label{eqn:coulombic}
\end{equation}
where $\chi_{C}$ is the Spitzer conductivity \citep{spitz62} and
$\boldsymbol{\hat{b}}$ is a unit vector in the direction of the
magnetic field.

We follow the HBI into the nonlinear regime using the conservative MHD
code Athena \citep{gs05} with the addition of anisotropic conduction
along magnetic field lines as described in \citet{ps05}.  This
methodology has been extensively utilized and tested previously in the
study of the MTI.  In all simulations, $g = 1$.  The 2D simulations use
a linear temperature gradient with fixed temperature boundary
conditions at the top and bottom of the domain.
\vspace{-0.05in}
\subsection{Physics of the HBI} \label{subsec:physics}
\begin{figure}[t!]
\epsscale{0.7}
\centering
\includegraphics[clip=true, scale=0.55]{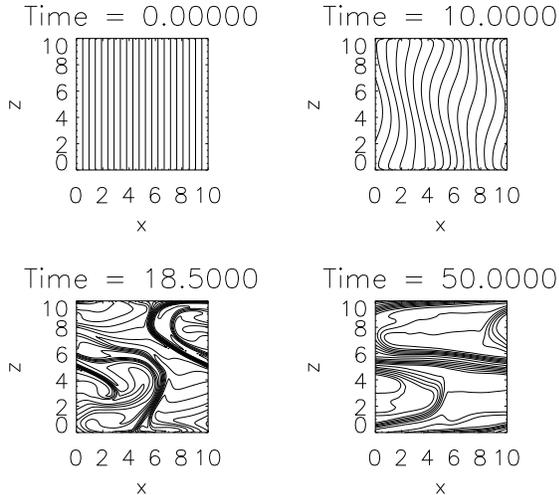}
\caption{Snapshots of the magnetic field in a 2D simulation
initialized with a single mode perturbation having $k_x=k_y$.  The HBI
drives the initially vertical field to become largely horizontal.  The
units of time in Figs. 1-4 are such that the dynamical time $(g
\dif\ln T/\dif z)^{-1/2} \approx 1.4$.}
\label{fig:2D-B}
\end{figure}

\vspace{-0.04in}
The physical origin of the HBI can best be understood by examining
magnetic field snapshots from a 2D simulation, as shown in Figure
\ref{fig:2D-B} (see also Fig. 1 of \citealt{q07}).
This 2D simulation was run at a resolution of (100)$^2$.  The initial
temperature profile was $T(z) = T_0(1+z/2)$, the temperature
scale-height at the midplane is $2$, and the size of the domain is
$(0.1)^2$; note that the simulations are local.  The pressure and
density were chosen so that the atmosphere was in hydrostatic
equilibrium with $p\sim \rho\sim 1$.  The
magnetic field is chosen to be weak initially and purely vertical with
$B_0/(4\pi)^{1/2}= 5\times 10^{-4}$, so that magnetic tension forces are
negligible. The purely anisotropic thermal diffusivity is $\kappa =
\chi_C T/P = 10^{-2}$, where $\kappa$ has units of a diffusion
coefficient, i.e., cm$^2$ s$^{-1}$.  With these parameters, the
conduction time for small-scale perturbations is much less than the
dynamical time, which is the limit of fastest growth for the HBI (and
MTI).

Figure \ref{fig:2D-B} shows that perturbations with non-zero $k_x$ and
$k_y$ generate converging and diverging field lines.  The heat flux
follows these field lines leading to (conductive) heating and cooling
of the plasma.  In a plasma with $\dif T/\dif z > 0$, a fluid element
displaced upwards is thus heated by the background heat flux, causing
it to rise further and become buoyant.  
Note that the magnetic field snapshots in Figure \ref{fig:2D-B} look
very similar to snapshots for the MTI rotated by 90$\degree$.

For a weak vertical magnetic field, the growth rate of the HBI in the
limit of rapid conduction is
\begin{equation}
\omega^2 \approx -g\left(\frac{\dif \ln T}{\dif
z}\right)\frac{k_{\perp}^2}{k^2}.
\label{eqn:gr}
\end{equation}
For the parameters of our simulation, we predict a growth rate of 0.5 for $k_x = k_y$.
Measurement of the growth rate in our single mode simulation verifies
this prediction to within 1\%.  In the 2D simulation, we find that the
magnetic energy is amplified by a factor of $\sim52$ during the course of
the run.  We defer a detailed discussion of the nonlinear saturation
to the 3D simulation.
\vspace{-0.156in}
\section{Nonlinear Saturation in 3D} \label{sec:3D}
\vspace{-0.09in}
Only in 3D can we accurately explore the saturation of the HBI since
the nature of convection differs significantly in 2D and 3D and the
anti-dynamo theorem applies in 2D \citep{cowl34}.  Using 3D
simulations, we now quantify the amplification of the magnetic field
by the HBI and the resulting heat flux through the plasma.

\begin{figure}[htb]
\epsscale{0.45}
\centering
\includegraphics[clip=true, scale=0.45]{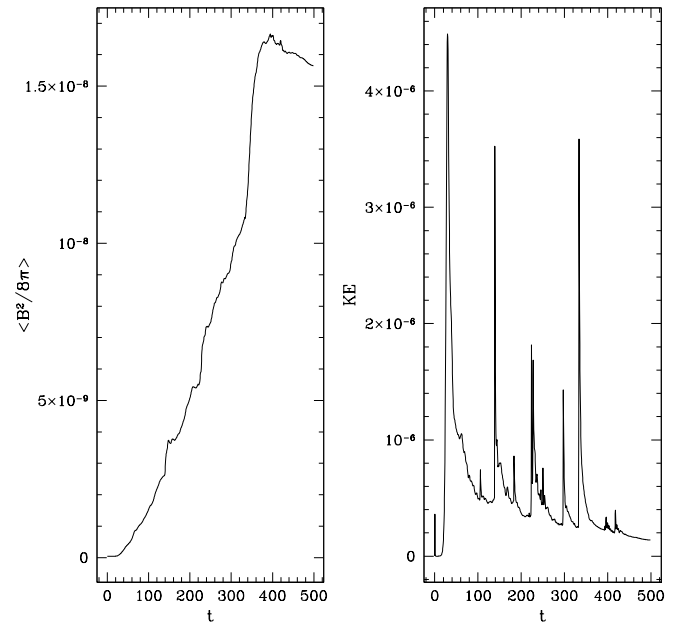}
\caption{Time evolution of the volume-averaged magnetic energy density
and kinetic energy density in the HBI-unstable region of run R1.  The
magnetic energy is amplified by a factor of $\sim 300$ and is in rough equipartition with the subsonic convective motions.}
\label{fig:fid-time}
\end{figure}
The computational set-up we use is nearly identical to the stratified
box presented in \S2.3 of \citet{ps07}; we refer the reader there for
details.  The box consists of three vertical layers of equal size in
which the central region has a linear temperature profile with
$\partial T/\partial z > 0$ and pure anisotropic conduction along
magnetic field lines, such that it is unstable to the HBI.  The top
and bottom layers are isothermal atmospheres (exponential pressure and
density profiles) and act as a buffer to the penetrative convection
that takes place.  These regions have isotropic
conductivity and are stable to the HBI. For all of the simulations in
this paper, the isotropic conductivity in the top and bottom layers is
equal to the parallel anisotropic conductivity in the middle region.
It is important to note that our box size is again small compared to
the temperature scale height (0.2 and 2, respectively), so that our
simulations are local in nature.  Table \ref{tab:runs} gives the
initial magnetic field and conductivity for our 3D runs; we will
primarily focus on R1, our fiducial simulation.

\begin{deluxetable*}{lcccccc}
\tablecolumns{7}
\tablecaption{Table of Nonlinear Runs  \label{tab:runs}}
\tablewidth{0pt}
\tablehead{
\colhead{Run} & 
\colhead{$B_0/(4\pi)^{1/2}$} &
\colhead{$\kappa_{\parallel}$} &
\colhead{$\Delta \la B^2 \ra$} &
\colhead{RMS Mach} &
\colhead{$\la \theta_B \ra$} &
\colhead{$f=Q/\tilde{Q}$}
}
\startdata
R1........ & $10^{-5}$ & $5\times 10^{-3}$ & 327 & $5.1\times 10^{-4}$ & $5.8\degree$ & 0.51\% \\
R2........ & $10^{-4}$ & $5\times 10^{-3}$ & 36.4 & $7.7\times 10^{-4}$ & $17.2\degree$ & 7.7\% \\
R3........ & $10^{-5}$ & $1.5\times 10^{-2}$ & 384 & $5.1\times 10^{-4}$ & $3.8\degree$ & 0.31\% \\
\enddata
\end{deluxetable*}

Figure \ref{fig:fid-time} shows the amplification of magnetic energy
as a function of time for R1.  The net amplification can be defined by
\begin{equation}
\Delta\langle B^2 \rangle \equiv \frac{\langle B^2 \rangle_{\textrm{fin}}}{\langle B^2 \rangle_{\textrm{init}}},
\label{eqn:bamp}
\end{equation}
which for the fiducial case is $\Delta \la B^2 \ra\sim 327$.  In
addition to amplifying the field, the HBI drives vertical convective
motions.  These remain subsonic with an RMS Mach number of $5.1\times
10^{-4}$; the kinetic energy is $\sim 10$ times the magnetic energy at the end of R1.

\begin{figure}[htb]
\epsscale{0.45}
\centering
\includegraphics[clip=true, scale=0.4]{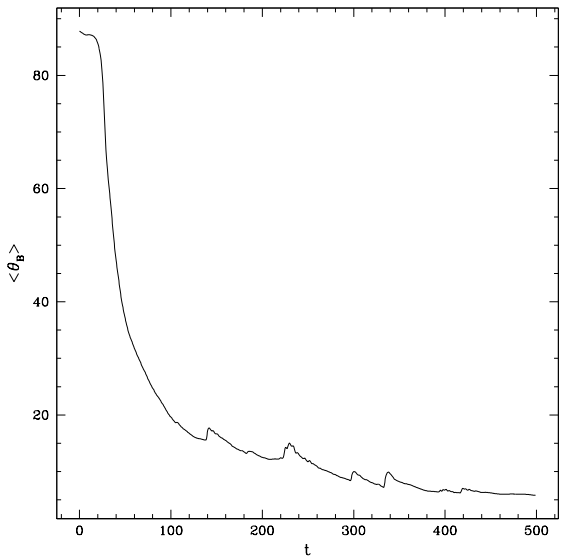}
\caption{Evolution of the average angle of the magnetic field with
respect to the horizontal for the fiducial simulation R1. The HBI
drives the initially vertical field to become almost purely horizontal
with an average angle of only 5.8$\degree$.}\label{fig:fid-btheta}
\end{figure}

\begin{figure}[htb!]
\epsscale{0.45}
\centering
\includegraphics[clip=true, scale=0.4]{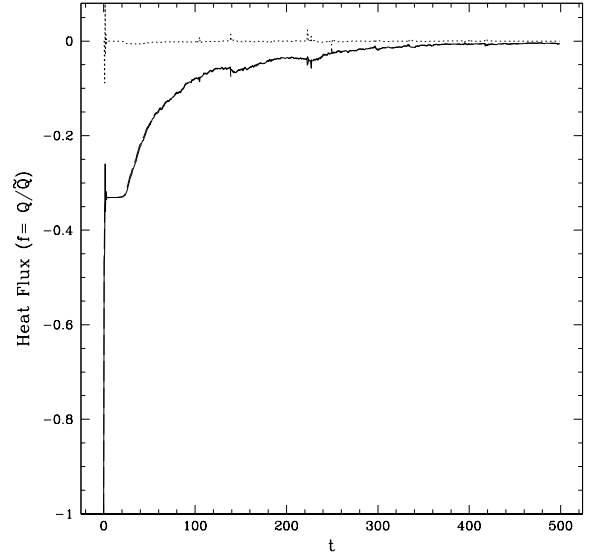}
\caption{Time evolution of the horizontally-averaged vertical heat
flux at the midplane in the fiducial simulation R1.  The total heat
flux (solid line) is divided into the dominant conductive flux (dashed
line) and the advective flux (dotted line). Note that the total heat
flux and conductive flux are essentially collinear.  These fluxes are
normalized to the fiducial heat flux, $\tilde{Q}$, that would be
present for vertical magnetic field lines
(eq. [\ref{eqn:Qfid}]).}\label{fig:fid-heat}
\end{figure}

The evolution of the average angle of the magnetic field with respect
to the horizontal in the unstable layer is particularly interesting;
this is defined by
\begin{equation}
\langle\theta_B\rangle \equiv \left\langle \sin^{-1} \left(\frac{|B_z|}{|B|}\right)\right\rangle.
\label{eqn:thetab}
\end{equation}
Figure \ref{fig:fid-btheta} shows $\langle \theta_B \rangle$ for run
R1.
The buoyant motions of the HBI saturate by driving the magnetic field
to be largely \emph{perpendicular} to the background temperature
gradient.  This evolution is in stark contrast to the MTI, which
preferentially aligns the magnetic field {\it with} the background
temperature gradient.  For R1, the magnetic field evolves from an
initially vertical field (90$\degree$) to a field with an average
angle of $\la \theta_B\ra \approx 5.8\degree$.  As one might expect,
this dramatically decreases the heat flux through the plasma, as is
shown explicitly in Figure \ref{fig:fid-heat}.
It is useful to compare the heat flux in the saturated state of the
HBI with the heat flux that is present in the initial state,
\begin{equation}
\tilde{Q} = - \kappa_{\parallel}\left(\frac{\dif T}{\dif z}\right)_{0}.
\label{eqn:Qfid}
\end{equation}
As the HBI evolves and the field lines become more horizontal
(Fig. \ref{fig:fid-btheta}), the heat flux across the unstable layer
plummets.  We define the conduction efficiency by $f = Q/\tilde{Q}$.
The saturated state of the HBI has a very low heat flux of
$f=5.0\times 10^{-3}$ for R1; most of this flux is carried by
conduction, with only a small contribution from convective motions
(Fig. \ref{fig:fid-heat}).  We will discuss the obvious implication of
this result for cooling flows shortly.

For a weak initial magnetic field, the saturation of the HBI is
primarily due to the reduction in the mean angle of the magnetic field
$\langle \theta_B \rangle$.
\citet{q07} showed that for nearly horizontal fields the maximum
growth rate of the HBI is given by
\begin{equation}
|\omega| \simeq (g\,\dif\ln T/\dif z)^{1/2} 
\la\theta_B\ra.
\label{eqn:gr2}
\end{equation}
At the conclusion of run R1, the growth rate has decreased by a factor
of $\sim 10$, to $\omega \simeq 0.07$, significantly reducing the
efficacy of the instability.  Meanwhile the growth of the magnetic
field increases the effects of magnetic tension. The Alfv\'en
frequency (a lengthscale dependent quantity) ranges from 0.01 to 0.7
from the size of the domain down to the grid scale, respectively.
Because the Alfv\'en frequency is comparable to the dynamical
frequency on small-scales, and to the growth rate on all but the
largest scales, tension will further inhibit the growth of the HBI.
Although we cannot rule out modest additional decrease in $\langle
\theta_B \rangle$ and the heat flux on timescales longer than we have
simulated, such evolution will take progressively longer as the growth
rate decreases further (eq. [\ref{eqn:gr2}]) and is unlikely to modify
the magnetic and kinetic energy, which have saturated
(Fig. \ref{fig:fid-time}).

For completeness we ran several additional simulations.  Run R3 has a
conductivity three times higher than the fiducial value.  However,
since R1 is already in the limit in which the conduction time for
small-scale perturbations is rapid compared to the dynamical time,
little change is seen from the fiducial simulation.  Run R2 has an
initial magnetic field one order of magnitude larger than the fiducial
run.  Due to magnetic tension effects on the scale of this local
simulation, R2 saturates earlier and with less magnetic field
amplification.  The magnetic field is still primarily horizontal upon
saturation and the heat flux is strongly suppressed, although by less
than in the fiducial simulation.  In a global simulation, the
saturated state of R2 would be unstable on large-scales, which would
lead to additional field amplification and rearrangement.
\vspace{-0.11in}
\section{Applications and Discussion} \label{sec:conc}
\vspace{-0.07in}
The HBI is predicted to be present from $\sim 0.1-100$ kpc in the
intracluster medium of galaxy clusters, where the observed temperature
increases outwards.  For concreteness, consider the cluster A1795
which has an estimated virial mass of $1.2\times 10^{15}$ M$_{\odot}$
and a scale radius of $r_s = 460$ kpc \citep{ettori02,zn03}.  The
temperature increases from $\simeq 2$ keV at $10$ kpc to $\simeq 7.5$
keV at 100 kpc.
Given these parameters, the growth rate of the HBI at $R = 50$ kpc in
the limit of weak fields is $\simeq 3 \times 10^{-16}$ s$^{-1}$, i.e.,
a growth time of $\simeq 10^8$ yr.  This should be compared to the
wavelength($\lambda$)-dependent Alfv\'en frequency of $\simeq 5 \times
10^{-17} \, (\lambda/R)^{-1} (B/{\rm 1 \mu G})$ s$^{-1}$ given the
measured electron density of $n_e \simeq 0.02$ cm$^{-3}$ at 50 kpc and
an assumed $B \sim 1 \mu$G.  Thus tension has a small effect for
$\lambda \gtrsim 0.1 R$.  Because the electron mean free path is $\sim
0.05 R$, the thermal conductivity is sufficiently high for
perturbations to grow on the dynamical time (eq. [\ref{eqn:gr}]).

Our simulations presented in \S3 were local ($L_z < H$), but their
nonlinear outcomes provide a clear lesson for galaxy clusters.  So
long as magnetic tension is not dominant, the HBI rapidly and
efficiently re-orients the magnetic field lines to be perpendicular to
the background heat flux, resulting in a net heat flux through the
plasma that is much less than the field-free Spitzer value.
Our results are significantly different from the predictions of
\citet{nm01}, \citet{cm04}, and \citet{laz06}, all of whom highlight
the effects of field line wandering on the effective thermal
conductivity of a plasma.  Instead, however, our results, and those of
\citet{bal00}, \citet{ps07}, \& \citet{q07}, demonstrate that the
dynamical coupling between anisotropic thermal conduction, the heat
flux, and the magnetic field geometry is crucial.  For HBI-unstable
plasmas, the net result is that heat is transported far less
effectively than previously assumed.  These conclusions make the
cooling flow problem in clusters more severe since the HBI strongly
suppresses the transport of energy from the outer parts of clusters
($r>100$ kpc) to the cooling core.  An alternate mechanism is thus
necessary to heat the cluster plasma, e.g., feedback from a central
AGN.  Several additional effects should be accounted for before this
conclusion is considered definitively established.  First, cosmic rays
from a central AGN modify the stability properties of intracluster
plasma \citep{cd06}.  The interplay between the cosmic rays, which
tend to drive the system MTI unstable, and the plasma, which by itself
is HBI unstable at intermediate radii, may be interesting and complex.
In addition, turbulence produced by galactic outflows and galaxy wakes
could modify the magnetic field geometry established by the HBI.
Finally, given that tension restricts the HBI to scales $\gtrsim 0.1
R$, global simulations are clearly required.

An additional potential application of the HBI in clusters is to the
suppression of conduction across cold fronts.  The canonical cold
front is found in the cluster A3667 \citep{vik01a, vik01b}; our
numerical values are drawn from this example.
%
The cold front is an abrubt change in plasma properties, with high
temperature, low density, low pressure plasma lying outside low
temperature, high density, high pressure plasma.  The temperature
gradient is thus opposite to both the pressure gradient and the
cluster's gravity---an HBI unstable situation. For the gradient scale
length, we take the 5 kpc upper limit from {\it Chandra} measurements
and estimate an HBI growth rate of $\omega \approx 3.8 \times
10^{-15}$ s$^{-1}$, corresponding to a growth time of 8.5 Myr.  For a
1 $\mu$G field, magnetic tension is modest for perturbations with
wavelengths of 5 kpc, suggesting that the HBI can indeed grow.
Although the HBI cannot account for the origin of cold fronts, it may
help account for their survival.  Namely, the HBI will drive the
magnetic field to be horizontal in the cold front, suppressing heat
transport across the cold front and thus maintaining the temperature
gradient in spite of the nominally short conduction times.

We thank Prateek Sharma for useful conversations.  EQ was supported in
part by NASA grant NNG06GI68G and the David \& Lucile Packard
Foundation.  IJP is supported by NASA through a Chandra Postdoctoral
Fellowship grant PF7-80049 awarded by the Chandra X-Ray Center, which
is operated by the Smithsonian Astrophysical Observatory for NASA
under contract NAS8-03060.

\end{document}